\documentclass[preprint,showpacs,preprintnumbers,amsmath,amssymb]{revtex4}

\usepackage{graphicx}
\usepackage{dcolumn}
\usepackage{bm}

\begin{document}

\title{Microwave Spectrometry for the Evaluation of the Dimensions of Coronary Stents}

\author{Gianluca~Arauz-Garofalo}
 \email{gian.arauz@ubxlab.com}
\author{V\'{i}ctor~L\'{o}pez-Dom\'{i}nguez}

\affiliation{Grup de Magnetisme, Departament de F\'{i}sica Fonamental, Facultat de F\'{i}sica, Universitat de
Barcelona, c. Mart\'{i} i Franqu\`{e}s 1, planta 4, edifici nou, 08028 Barcelona, Spain}

\author{Oriol Rodriguez-Leor}
\author{Antoni Bayes-Genis}

\affiliation{Servei de Cardiologia, Hospital Universitari Germans Trias i Pujol, Carretera del Canyet s/n, ES-08916 Badalona (Spain)}

\author{Juan M.~O'Callaghan}

\affiliation{Department of Signal Theory and Communications, Universitat Politècnica de Catalunya, Jordi Girona 1, ES-08034 Barcelona (Spain)}

\author{Antoni Garc\'{i}a-Santiago}
 \email{agarciasan@ub.edu}
\author{Javier Tejada}

\affiliation{Grup de Magnetisme, Departament de F\'{i}sica Fonamental, Facultat de F\'{i}sica, Universitat de
Barcelona, c. Mart\'{i} i Franqu\`{e}s 1, pta. 4, edifici nou, 08028 Barcelona, Spain}

\date{\today}

\begin{abstract}
We study microwave scattering spectra of metallic stents in open air. We show that they behave like dipole antennas in terms of microwave scattering and they exhibit characteristic resonant frequencies for a given nominal size. We obtain a fair agreement between measured frequencies and the values provided by a theoretical model for dipole antennas. This fact opens the door to obtaining methods to detect structural distortions of stents within \textit{in vitro} conditions. Finally we discuss the \textit{in vivo} applicability of the suggested method in terms of our theoretical model and the skin depth of microwaves in biological tissues.

\end{abstract}


\maketitle

\section{INTRODUCTION}

A coronary stent \cite{Garg10, Garg10bis} is a medical prosthetic device shaped as a small cylindrical tube with wire mesh walls, which is used to rehabilitate atherosclerotic stenosed coronary arteries. The first generation of coronary stents, conventional bare-metal stents, was developed in the mid-1980s \cite{Schatz87}. Despite their obvious advantages in-stent neointimal hyperplasia \cite{Karas92, Gordon93, Hoffmann96} occurred in some cases. This phenomenon, also known as in-stent restenosis, was directly linked to stent implantation and resulted in restenosis rates of $20\%$ to $30\%$ \cite{Moliterno05}. It was the attempt to minimize this problem, and thereby reduce rates of repeat revascularization, that ultimately lead to the development of another revolutionary treatment: the drug-eluting stents. The dramatic reduction in restenosis rates seen with the use of these drug-eluting stents compared with bare-metal stents \cite{Stettler07, Spaulding07, Stone07, Mauri07, Kastrati07} has been the major driving force behind the exponential growth of percutaneous coronary interventions as a treatment for patients with coronary artery disease.

However, even with such improvements, patients with coronary stents require chronic medication and monitoring. For this reason considerable effort has been made to improve their quality of life. On the one hand, research has been carried out to improve the design of stents. Promising examples are biodegradable stents \cite{Ormiston09} or devices allowing wireless monitoring of pressure and blood flow \cite{Chow10, Takahata06}. Manufacturing methods based on microelectrodischarge machining have also been suggested \cite{Takahata04}. On the other hand, work is being conducted for the optimization of medical imaging techniques aiming for real-time high-quality images with minimal damage to the patient. These include non-invasive techniques such as magnetic resonance \cite{Eggebrecht06} and duplex ultrasound \cite{Wetzner84} as well as invasive methods such as X-ray angiography \cite{Perrenot07}, intravascular ultrasound \cite{Nissen01}, intravascular photoacoustic imaging \cite{Wang10} and optical coherence tomography \cite{Kauffmann10}. In spite of the obvious clinical benefits, these techniques require high costs and justification of potential patient collateral damages, like impact of invasive procedures or ionizing radiation dose.

Physicians have widely studied the medical risks arising from the structural distortion of stents. In particular, stent recoil \cite{Tsunoda04} and fracture \cite{Chakravarty10, Canan10} involve respectively drastic changes in diameter and length that could imply serious consequences. We show here that stents exhibit characteristic resonant frequencies in their microwave absorbance spectra which provide relevant information regarding their dimensions and should therefore reflect the occurrence of such structural distortions.

\begin{figure}
	\includegraphics[width=8.5cm]{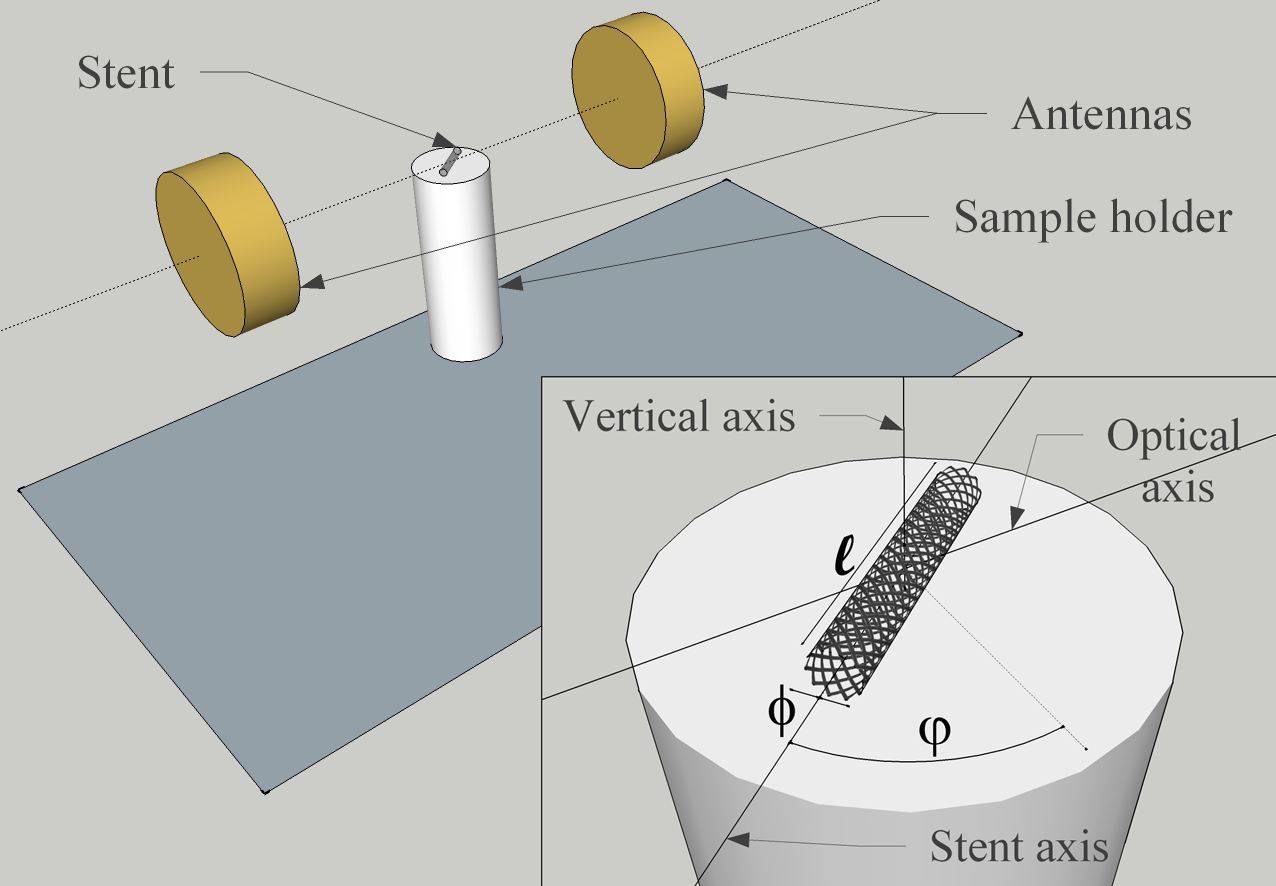}
	\caption{(Color online). Sketch of the experimental set-up. The inset shows a detail of the geometrical configuration for the characterization of stents.}
	\label{Fig1}
\end{figure}

\section{EXPERIMENTAL METHODS}

Figure \ref{Fig1} shows a sketch of the experimental set-up used in the present work. \textit{In vitro} microwave absorbance spectra were measured in open air for thirty commercial drug-eluting stents \cite{Medtronic} of different nominal length $\ell$ and diameter $\phi$. To obtain the spectra a pair of right-handed circularly-polarized cavity-backed spiral antennas \cite{A-info} was connected to a 2-port vector network analyzer \cite{HP} via coaxial feed lines. The antennas have a $2.0$ - $18.0$ GHz spectral range. This set-up provides the transmission coefficient of port 1 to port 2 as a function of frequency, $S_{12}\left( f \right)$, or vice versa $S_{21}\left( f \right)$, from which we can calculate the absorbance spectrum as
\begin{equation}
 \label{Eq1}
 A \left( f \right) =
 10 \log \left( \frac{S_{ref}\left( f \right)}{S\left( f \right)} \right),
\end{equation}
where $S\left( f \right)$ and $S_{ref}\left( f \right)$ respectively denote the transmission coefficients measured as a function of frequency with and without a stent, and subindices have been dropped for simplicity. Results were obtained from the average of 100 acquisitions of both magnitudes. Measurements were performed in a symmetrical configuration in which the center of the stent was placed at the midpoint of the line joining the two antennas (optical axis), which are set at a constant separation of 16 cm. The stents were positioned with the aid of an expanded polystyrene sample holder which allows to rotate the stent longitudinal axis (stent axis) at an azimuth angle $\varphi$ around a vertical axis that is perpendicular to the optical axis (see the inset in Figure \ref{Fig1}). $A \left( f \right)$ spectra were obtained at 4$^\circ$ steps over a complete rotation of each stent around the vertical axis.
\begin{figure}
	\includegraphics[width=8.5cm]{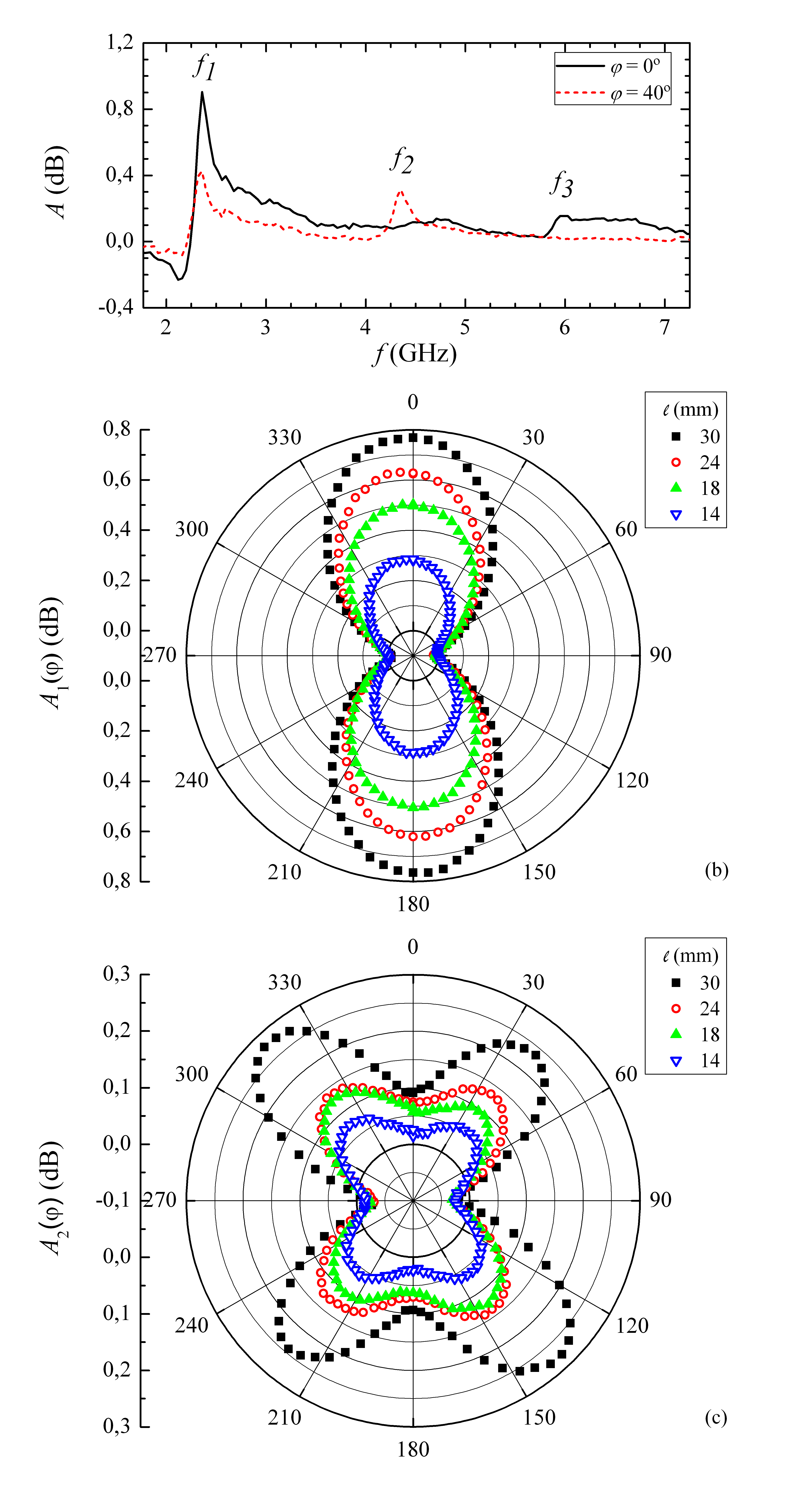}
	\caption{(Color online) (a) Absorbance spectra, $A\left( f \right)$, of a drug-eluting stent ($2.50$ mm in diameter and $30$ mm in length) at 0$^\circ$ and 40$^\circ$ values of $\varphi$ (solid and dotted line, respectively). (b) and (c) Angular dependence of absorbance for the first two resonances, $A_1\left(\varphi\right)$ and $A_2\left(\varphi\right)$, respectively, of four drug-eluting stents with the same diameter ($2.50$ mm) and different lengths from $14$ to $30$ mm (see legend for symbol details).}
 \label{Fig2}
\end{figure}
\begin{figure}
  \includegraphics[width=8.5cm]{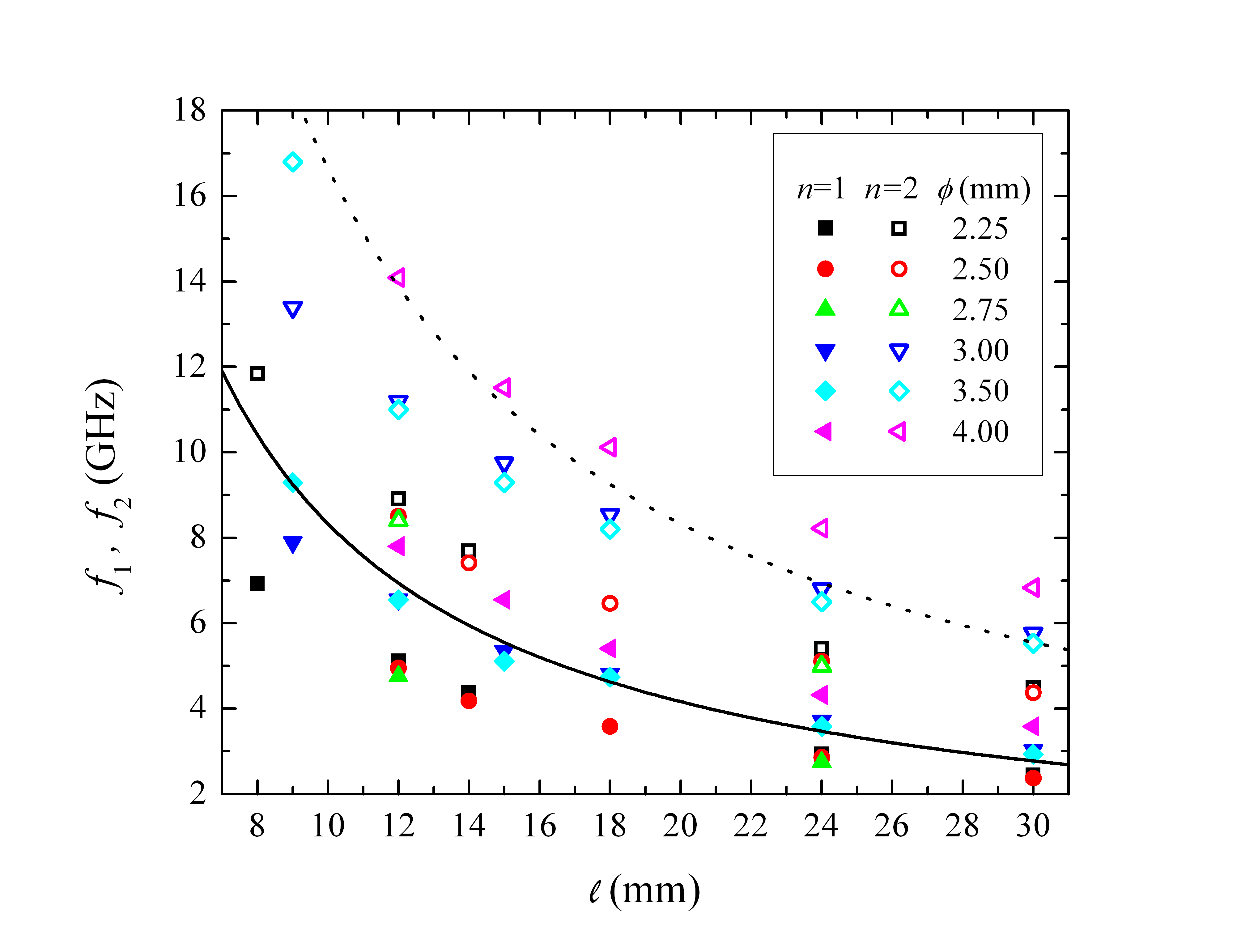}
	\caption{(Color online) Measured characteristic resonant frequencies as a function of length $\ell$ for several diameters $\phi$ of the thirty drug-eluting stents investigated. Solid and open symbols correspond respectively to the first $(f_1)$ and second $(f_2)$ experimental modes. The solid and dotted lines represent respectively the first and second theoretical modes given by Equation \ref{Eq2} for a scaling factor $a = 1.8$.}
	\label{Fig3}
\end{figure}

\section{RESULTS AND DISCUSSION}

Figure \ref{Fig2}(a) shows typical $A \left( f \right)$ spectra obtained at two values of $\varphi$ for a stent of well-defined dimensions. Similar spectra were obtained at any $\varphi$ value for the thirty stents investigated. Discrete resonant frequencies with quality factors close to 20 are exhibited by both spectra. Although up to five orders of resonance were detected, we focus here on the characterization of the first two, $f_1$ and $f_2$. All first orders on the one hand, and all second orders on the other, shared the same features for each studied stent. Figures \ref{Fig2}(b) and \ref{Fig2}(c) present respectively the azimuthal dependence of the amplitude of the first two resonances, $A_1\left(\varphi\right)$ and $A_2\left(\varphi\right)$, for four stents with the same diameter and different lengths. $A_1\left(\varphi\right)$ shows a bilobular pattern, with peaks occurring when the stent axis is perpendicular to the optical axis, while $A_2\left(\varphi\right)$ shows a tetralobular pattern, with peaks occurring when $\varphi$ is in the range 32$^\circ$ - 40$^\circ$. Higher resonances show more complex $A \left( \varphi \right)$ dependences, like hexalobular patterns for $f_3$, for example. The ability to discern these resonances by means of the inspection of their $A\left(\varphi\right)$ patterns allows to create a database with their values for each stent. Figure \ref{Fig3} shows $f_1$ and $f_2$ as a function of $\ell$ for several $\phi$ values of the thirty stents investigated. Notice how both frequencies are inversely proportional to $\ell$ and roughly proportional to $\phi$.

The shapes of the experimental curves in Figures \ref{Fig2}(b) and \ref{Fig2}(c) closely resemble gain patterns of a center-feed half-wave dipole antenna \cite{Orfanidis08}. This fact, along with the $1/\ell$ dependence shown in Figure \ref{Fig3}, suggests that the microwave scattering of stents is remarkably similar to that produced by such device \cite{PreNoteChow09}. The microwave electromagnetic field couples here to the antenna conductive structure and induces a standing electric current along it. Consequently the scattering is enhanced at resonant frequencies given by \cite{Balanis97}
\begin{equation}
 \label{Eq2}
 f_n (L) = \frac{n}{2 L \sqrt{\epsilon \mu}},
\end{equation}
where $\epsilon$ and $\mu$ are respectively the permittivity and permeability of the stent surrounding medium, $L$ is the dipole antenna length, and $n$ is the resonance mode. We found out that the resonant frequencies in a stent of length $\ell$ are significantly lower than those corresponding to a dipole antenna with $L = \ell$. This means that, in terms of microwave scattering, metallic stents of length $\ell$ behave akin to dipole antennas of length $L = a \ell$, where $a$ is a scaling factor. As an example, in Figure \ref{Fig3} we have plotted Equation \ref{Eq2} for $a = 1.8$. There is an acceptable agreement between experimental resonant frequencies and theoretical estimations. The presence of a scaling factor greater than one should be considered reasonable due to the characteristic folded structure of coronary stents. The precise value of this scaling factor would actually give a hint about the folding degree of a particular stent architecture.

\begin{figure}
  \includegraphics[width=8.5cm]{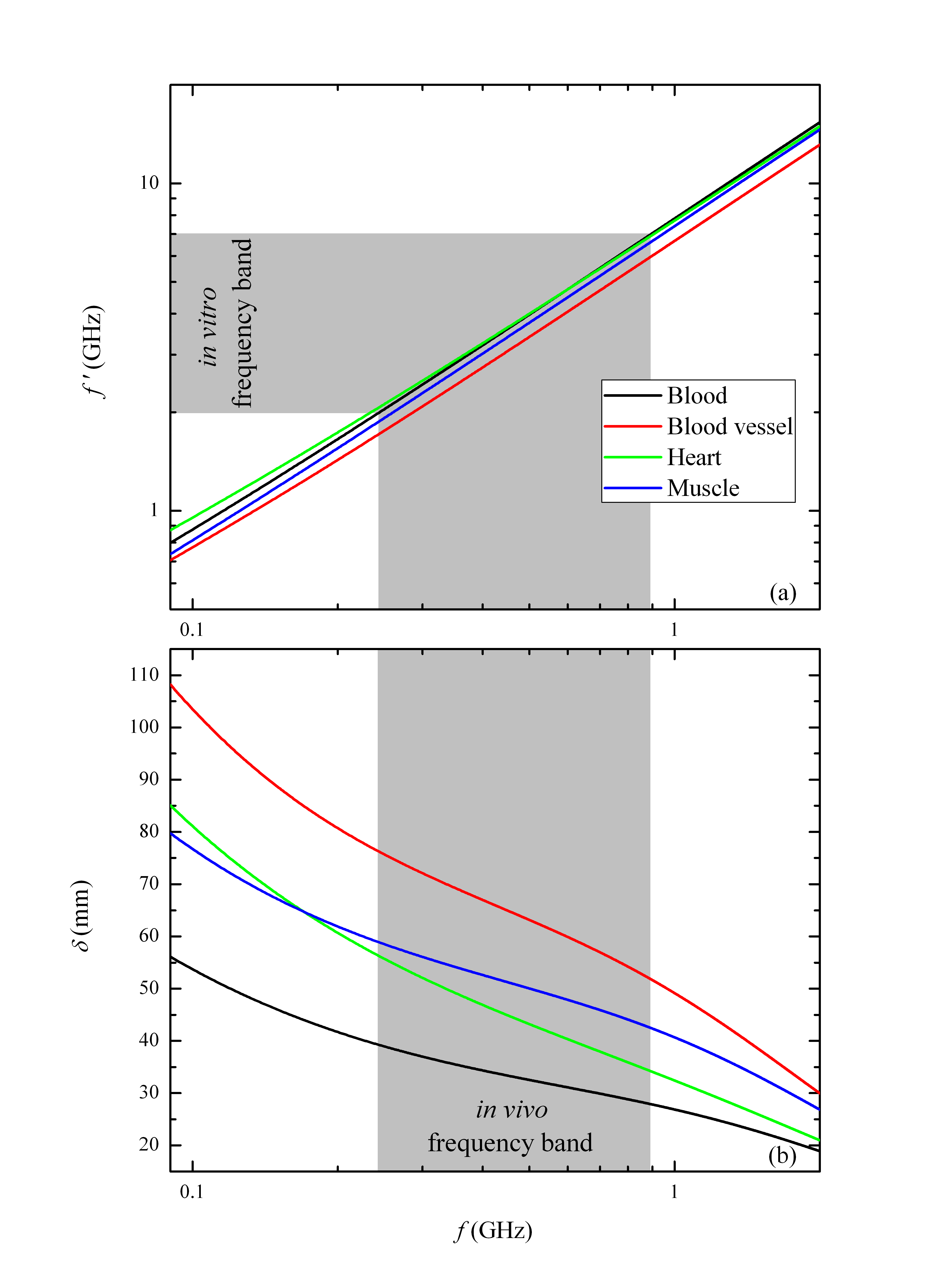}
	\caption{(Color online) (a) Relation between the \textit{in vitro} and \textit{in vivo} resonant frequencies, $f'(f)$, of the studied stents for four representative human biological tissues (see legend for details). (b) Skin depth as a function of frequency, $\delta(f)$, for the same four tissues. The shaded area highlights the range of $f$ (\textit{in vivo} frequency band) associated to the typical range of $f'$ (\textit{in vitro} frequency band) enclosing the first mode resonant frequencies of the measured stents.}
	\label{Fig4}
\end{figure}

These results show that a method for the detection of structural distortions of stents by measuring their resonant frequencies is possible, at least under \textit{in vitro} conditions. We will then discuss the \textit{in vivo} applicability of this method in terms of the skin depth of microwaves in the human body. Due to the $1/\sqrt{\epsilon \mu}$ factor of Equation \ref{Eq2}, it is expected that the \textit{in vitro} resonant frequencies of stents will shift down with respect to the \textit{in vivo} frequencies, since the relative permittivity of biological tissues in the microwave range is higher than one. We can thus obtain the relation between the \textit{in vitro} and \textit{in vivo} values, $f'$ and $f$ respectively, using Equation \ref{Eq2} and assuming air as the surrounding \textit{in vitro} medium of the stent,
\begin{equation}
 \label{Eq3}
 f'(f) =
 c f \sqrt{\epsilon(f) \mu(f)}.
\end{equation}
%where $\epsilon_r(f)$ and $\mu_r(f)$ are respectively the relative permittivity and permeability of the \textit{in vivo} biological tissue that surrounds the stent.
Introducing the dependences of $\epsilon(f)$ and $\mu(f)$ given by parametric models \cite{Gabriel96III} in Equation \ref{Eq3}, $f'(f)$ may be determined numerically. Figure \ref{Fig4}(a) shows such dependence for several representative tissues and highlights that the shift between $f'$ and $f$ is of about one order of magnitude.

Conduction and displacement currents induced by electromagnetic waves are of the same order for most biological tissues \cite{VanderVorst06}, so in this case the skin depth for media limited by plane boundaries is generally expressed as \cite{Jordan50}
\begin{equation}
 \label{Eq4}
 \delta(f) = \frac{1}{2 \pi f}\left\{\frac{1}{2} \mu(f) \epsilon(f) \left( \sqrt{1 + \left( \frac{\sigma(f)}{2 \pi f \epsilon(f)}\right)^2} - 1 \right)\right\}^{-1/2},
\end{equation}
where $\sigma$ is the conductivity of the medium. Using again parametric models \cite{Gabriel96III}, we can determine numerically $\delta(f)$. Figure \ref{Fig4}(b) shows that $\delta(f)$ is found to be of several tens of millimeters for \textit{in vivo} operating frequencies. This fact demonstrates that microwaves travel through the thoracic cavity and reach the heart, thereby enabling the medical applicability of microwave spectrometry for the evaluation of the dimensions of implanted coronary stents.

\section{CONCLUSIONS}

We have proved experimentally that metallic stents of a given nominal length and diameter exhibit characteristic resonant frequencies in their open air microwave scattering spectra. We have also proposed a simple theoretical model based on a dipole antenna that allows to estimate the values of such frequencies. It is expected that some structural distortions experienced by implanted stents due to aging should be reflected in a variation of the resonant frequencies. Finally, we have estimated the \textit{in vivo} frequency band corresponding to the \textit{in vitro} frequencies of the investigated stents and its associated skin depth range for several representative tissues, demonstrating the medical applicability of the method here discussed. Our work will allow the achievement of techniques able to prevent alterations, such as stent recoil or fracture, that nowadays have strong medical impact. Characterization of stents with induced geometrical distortions within biological tissue mimics are under development. Further experiments will include \textit{ex vivo} and \textit{in vivo} trials. We hope that our work will stimulate the development of non-invasive and non-ionizing intensive monitoring medical techniques of patients with coronary stents.

\section{ACKNOWLEDGMENTS}
G. A.-G. thanks L. Humbert-Vidan for her backing, and ICREA Acad\`{e}mia and Universitat de Barcelona for the financial support. V. L.-D. and J. T. appreciate financial support from ICREA Acad\`{e}mia and Universitat de Barcelona. A. G.-S. thanks Universitat de Barcelona for backing his research. O. R.-L. and A. B.-G. thanks Medtronic, Inc. for providing the stents studied in the present work. J. O'C. thanks the Spanish Government project MAT2011-29269-C03-02.

%\begin{thebibliography}
\bibliography{arauz garofalo 20120723}
%\end{thebibliography}

\end{document}